# BIAS, DIVERSITY, AND CHALLENGES TO FAIRNESS IN CLASSIFICATION AND AUTOMATED TEXT ANALYSIS. From libraries to AI and back

Bettina Berendt[2,3,1] *, Özgür Karadeniz[1], Sercan Kıyak[1], Stefan Mertens[1], Leen d'Haenens[1]

[1] KU Leuven, Belgium; [2] TU Berlin, Germany; [3] Weizenbaum Institute, Germany
* corresponding author <firstname>.<lastname>@kuleuven.be

8 March 2023

**Abstract:** Libraries are increasingly relying on computational methods, including methods from Artificial Intelligence (AI). This increasing usage raises concerns about the risks of AI that are currently broadly discussed in scientific literature, the media and law-making. In this article we investigate the risks surrounding bias and unfairness in AI usage in classification and automated text analysis within the context of library applications. We describe examples that show how the library community has been aware of such risks for a long time, and how it has developed and deployed countermeasures. We take a closer look at the notion of '(un)fairness' in relation to the notion of 'diversity', and we investigate a formalisation of diversity that models both inclusion and distribution. We argue that many of the unfairness problems of automated content analysis can also be regarded through the lens of diversity and the countermeasures taken to enhance diversity.

## 1 Introduction

Libraries are increasingly relying on computational methods, including methods from Artificial Intelligence (AI)[1]. This increasing usage raises concerns about the risks of AI that are currently broadly discussed in the scientific literature, the media and law-making. But libraries have also, for a long time, been at the forefront of innovative ways of dealing with information and leveraging knowledge-processing methods for the common good. In this article, we first investigate risks surrounding bias and unfairness that have recently come under increasing scrutiny as possible side-effects of AI (Section 2). We then ask, in Section 3, how such risks can also affect library applications, using automated content analysis as an application of machine-learned classification in which such risks have been observed. We also consider how the library community has been aware of such risks for a long time, and point at some countermeasures that have been developed and deployed by librarians.

---

[1] For examples, see the activities of the German/German-language *Netzwerk maschinelle Verfahren in der Erschliessung* (Network machine methods in indexing and classification), https://wiki.dnb.de/display/FNMVE/ or the research presented at the international *Semantic Web in Libraries Conference* at https://swib.org.

We start this analysis by taking a closer look at the notion of "(un)fairness", linking it to the notion of "diversity" by arguing that a key prerequisite for being treated fairly is that an individual or group has to "exist" (in representations) in order to be perceived. We investigate a formalisation/operationalisation of diversity that encompasses these two ideas by modelling both distribution and (prior to this) inclusion. We conclude, in Section 4, that many of the unfairness problems of automated content analysis can also be regarded through the lens of diversity and the countermeasures taken to enhance diversity – the visibility of and respect for diversity as it exists in the world, and its presence in content and metadata.

## 2 Key concepts

Within and across disciplines, '*fairness*' is a multi-faceted and contested concept.[2] In machine learning, fairness most often refers to some form of equality of measurable outcomes of algorithmic decision-making, where all components of the definition, including what exactly should be equal (or diverge by only a small quantity) are being disputed.[3] An important type of unfairness stems from biased representations that display and often perpetuate stereotypes against groups of people. Such representations can lead to or exacerbate discrimination against individuals and groups.

Algorithmic unfairness and *algorithmic discrimination (AD)* denote unfairness and discrimination in contexts that involve (usually digital) computers. Friedman and Nissenbaum argued, against a then frequent conception of computers as 'more objective' than humans, that computer systems can in fact be biased, which can lead to discrimination. They "use the term bias to refer to computer systems that systematically and unfairly discriminate against certain individuals or groups of individuals in favor of others".[4]

These groups of individuals are usually identified by categories, which often (but not always) coincide with "legally protected grounds" and which generally are socially normalised. Bourdieu argues that the categories through which social agents exist and are known are a matter of continuous political struggle over knowledge of the social world: "(T)his work of categorisation, i.e., of making explicit and of classification, is performed incessantly, at every moment of ordinary existence, in the struggles in which agents clash over the meaning of the social world and of their position within it."[5]

---

[2] Berendt, Bettina; Karadeniz, Özgür; Mertens, Stefan and d'Haenens, Leen: Fairness beyond "equal": The Diversity Searcher as a Tool to Detect and Enhance the Representation of Socio-political Actors in News Media, in: Companion Proceedings of the Web Conference 2021, Ljubljana Slovenia 2021, pp. 202–212. Online: <https://doi.org/10.1145/3442442.3452303>, Retrieved: 25.08.2022.

[3] Žliobaitė, Indrė: Measuring discrimination in algorithmic decision making, in: Data Mining and Knowledge Discovery 31 (4), 07.2017, pp. 1060–1089. Online: <https://doi.org/10.1007/s10618-017-0506-1>; Hutchinson, Ben; Mitchell, Margaret: 50 Years of Test (Un)fairness: Lessons for Machine Learning, in: Proceedings of the Conference on Fairness, Accountability, and Transparency, Atlanta GA USA 2019, pp. 49–58. Online: <https://doi.org/10.1145/3287560.3287600>

[4] Friedman, Batya; Nissenbaum, Helen: Bias in computer systems, in: ACM Transactions on Information Systems 14 (3), 07.1996, pp. 330–347. Online: <https://doi.org/10.1145/230538.230561>..

In this context, it is becoming increasingly clear that classifications of people into the "protected group" and the "unprotected group" (a classification inspired by unfairness viewed as unlawful discrimination) and the equalisation of some measures of outcomes, which dominate the AI fairness literature, can only capture some aspects of a wider notion of distributive justice.[6] In this article, we consider diversity as one of the important prerequisites of fairness. One reason for this relates to Bourdieu's observation: individuals and groups must first be visible (through a categorisation) to claim fair treatment.

*Diversity* has been investigated in a wide set of domains including ecology, the physical, social, life and information sciences, economic and policy studies, communication, and cultural analysis.[7] We draw on an influential proposal of a quantitative measure that is based on a meta-analysis of this wider literature. Andy Stirling identified three components, which all enhance diversity:

- **Variety**: many distinct entities are present, e.g., a number of people of different backgrounds;
- **Disparity**: the entities are different from one another, they come from a wide range;
- **Balance**: the different entities are evenly distributed. This component is a simple form of the "equality of ..." concept found in the fairness literature.[8]

Based on these three components, Stirling defined diversity Δ as

$$\Delta = \sum_{i,j} (d_{ij})^{\alpha} \cdot (p_i \cdot p_j)^{\beta}$$

with variety being captured by the cardinality of the set of all entities i, j ∈ E present in the domain (e.g., a text, a population, ...), balance by the frequencies $p_i$ (the more uniform the distribution, the higher this multiplicative factor), and disparity by a measure of distance or dissimilarity d(.,.). In addition, the parameters α and β allow for a weighting of the importance of balance or disparity.

---

[5] Bourdieu, Pierre: The Social Space and the Genesis of Groups, in: Theory and Society 14 (6). 1985, pp.723-744.
[6] Binns, Reuben: Fairness in Machine Learning: Lessons from Political Philosophy, in: Proceedings of the 1st Conference on Fairness, Accountability and Transparency, PMLR, (n.d.), pp. 149–159.
[7] Stirling, Andy: A general framework for analysing diversity in science, technology and society, in: Journal of The Royal Society Interface 4 (15), 22.08.2007, pp. 707–719. Online: <https://doi.org/10.1098/rsif.2007.0213>; Ranaivoson, Heritiana: Measuring cultural diversity with the Stirling model, in: New Techniques and Technologies for Statistics, 2013, p. 10. Online: <https://ec.europa.eu/eurostat/cros/content/measuring-cultural-diversity-stirling-model-heritiana-ranaivoson_en>.
[8] Stirling: A general framework for analysing diversity in science, technology and society, 2007.

This formula can measure various forms of diversity, including *actor diversity*. Actor diversity can be measured in a real-world setting (where one could ask, for example, how diverse a company's workforce is, based on the question "who is on this company's payroll?") or in a representation (where one could ask, for example, how diverse a text is, based on the question "who appears in this text?"). Ultimately, of course, the first kind of scenario also involves modelling reality and in this sense it also depends on representations. In the remainder of this paper, we will focus on actors in representations (such as texts or library-related metadata).

The diversity formula models one meaning of *inclusion*: variety measures who is present in the representation and included in it. This can also be regarded as the *visibility* of these actors. The formula also measures one meaning of *distribution*: balance measures how frequently different actors occur in the representation. These notions relate to mathematical sets and statistics.

For *fairness* to be measurable across distinct individuals or groups in data describing a distribution of these distinct entities, there must be diversity in this sense, more specifically actor diversity. Fairness is more than actor diversity, however, because it also describes an allocation of some other entity (goods, services, opportunities, …) across these individuals or groups, or because it describes a representation of these individuals or groups in which, for example, ascriptions of properties are "fair" to them (and possibly others) rather than "stereotyped" or "biased".

This understanding of fairness alludes to another meaning of *distribution*, related to distributive justice. In the algorithmic fairness literature, this is often referred to as the avoidance of allocative harms[9], and the "de-biasing" of data and algorithmic decision-making is proposed as a computational countermeasure. The algorithmic fairness literature also discusses representational harms[10] created for example by stereotypes in texts or data (and how to detect and avoid them). Both types of harms interfere with (and countermeasures seek to promote) *inclusion* in the sense that not only different actors are present, but that they can all flourish. These wider notions of inclusion and distribution relate to philosophical and political concepts rather than mathematics and statistics alone.

There can thus be diversity without fairness, and too much diversity could also harm fairness (cf. for example the long-standing discussion about how too much "neutrality" in the media can also push extremist and discriminatory standpoints and thus harm vulnerable groups).

In the present article, we focus on the positive side: Diversity as an enabler of fairness, and unbiasedness as an enabler of diversity. In Section 3, we study unfairness challenges and diversity-enhancing solution approaches in the library context.

---

[9] Blodgett, Su, Barocas, Solo, Daumé III, Hal and Wallach, Hanna, Language (Technology) is Power: A Critical Survey of "Bias" in NLP, in: Proceedings of the 58th Annual Meeting of the Association for Computational Linguistics, Stroudsburg / PA 2020, 5454-5476 B
[10] ibid.

# 3 Algorithmic unfairness: Challenges and solution approaches in library contexts

Algorithmic unfairness and algorithmic discrimination can result from interactions between data, algorithms and the larger socio-technical systems they are used in. In particular, algorithms that "learn" from biased data will produce biased outputs; algorithms themselves may encode biases, machine-learned classification may in itself be unfair, and chains and feedback loops of such effects will reproduce and exacerbate unfairness.[11] Here, "bias" may arise from different reasons including stereotypes explicitly baked into the data, from problems with sampling (sampling bias, representation bias), and it may be caused by constraints imposed by formalisations and technology or emerge during usage.[12] Computer scientists have developed a large number of methods for "de-biasing" data, algorithms, and results.[13] However, the aim of fully de-biasing them may be as difficult to achieve as to "stop discrimination" in general.

Such phenomena have long been observed in library contexts. In response, librarians and information scientists have developed and applied strategies for mitigating such effects. This section will illustrate this with examples that we map to an idealised linear pipeline that spans indexing, algorithms, ontologies, and user interactions. We will see that librarians and library scientists have been particularly concerned with representational harms. In this sense, they are pioneers in finding solutions to an issue that has only recently entered the mainstream of AI researchers' attention.

## 3.1 Challenges for automated content analysis through a lens of fairness

Algorithmic or algorithm-enhanced content analysis is practised by an increasing number of libraries.[14] For example, the German National Library's *Automatisches Erschließungssystem* project uses AI methods to learn from existing subject cataloguing. Unstructured texts and metadata of new materials are then mapped to subject categories taken, inter alia, from the Dewey Decimal Classification (DDC) and the Library of Congress Subject Headings (LCSH).[15]

---

[11] Berendt, Bettina: Algorithmic discrimination, in: Comandé, Giovanni (Ed.): Elgar encyclopedia of law and data science, Cheltenham, UK 2022.

[12] Hellström, Thomas; Dignum, Virginia; Bensch, Suna. Bias in Machine Learning - What is it Good for? In Proceedings of the First International Workshop on New Foundations for Human-Centered AI (NeHuAI) co-located with 24th European Conference on Artificial Intelligence (ECAI 2020). CEUR-WS Vol. 2659. 2020. Online: https://ceur-ws.org/Vol-2659/hellstrom.pdf, Retrieved: 06.03.2023; Friedman and Nissenbaum (1996), see footnote 4.

[13] See footnote 11.

[14] See for example the publications and workshops organised by the 'network machine methods in content indexing and classification', see footnote 1, or the *Automation of Subject Indexing using Methods from Artificial Intelligence* project at https://www.zbw.eu/en/about-zbw/key-activities/automation-subject-indexing

[15] Another key vocabulary is GND (Gemeinsame Normdatei, Integrated Authority File), which focusses on named entities. It is possible that bias and fairness issues arise also here, but this is a topic for another article.

This information architecture poses three challenges: existing subject indexing; the algorithms employed in learning and cataloguing; and the classification systems themselves. We will order these based on our perspective from AI. Much of the impact of classification systems arises when they are being used unwittingly and much impetus for change arises from individuals and groups who object to the unfairness they perceive in this and who create innovative work-arounds. These actions use and produce the "data" that are today strongly perceived as carrying bias. Algorithms are the second (and sometimes even the first) element in the spotlight when the 'risks of AI' are being discussed. A deeper look at the underlying categories and ontologies is still much rarer and – in the general case of AI – also much more difficult, since the socio-technical systems in which AI is being deployed are much harder to change than an algorithm or a dataset. We therefore treat existing subject indexing as challenge #1, algorithms as challenge #2, and classification systems as challenge #3.[16]

This information architecture leads to challenge #1: existing subject keywording (aka subject indexing). These mappings of contents to categories reflect a history of thought and thought practices, a history that can come with severe biases. We will first consider responses, often individual, to such keywording *as-is*. These responses need to be, and are, accompanied by moves to address issues of bias on the systemic side of the classification system per se, see challenge #3 below.

In her cataloguing work during the 1930s, Dorothy Porter, librarian at Howard University, a historically black college, consulted a large number of libraries for guidance that all relied on DDC. She reported: "Now in [that] system, they had one number—326—that meant slavery, and they had one other number—325, as I recall it—that meant colonization. In many 'white libraries,' every book, whether it was a book of poems by James Weldon Johnson, who everyone knew was a black poet, went under 325. And that was stupid to me." (Porter cited by Nunes, 2018).[17] Porter wrote history by re-assigning these works based on content criteria rather than author demographics.[18]

While the data that Porter observed would immediately strike most modern-day librarians or readers as racist, other arguably biased classifications can be observed today: "Books that cover the history of the civil rights movement, immigrant histories, and women's history were getting sent to the 300s [Social Science]", not to the 900s [History] (Jess de Courcy Hines, reported by Schwartz, 2018).[19]

---

[16] We have received a comment from a librarian that from their perspective, the reverse order would make more sense. We consider this a very interesting comment; to us it suggests that the library community may consider themselves strongly in the role of defining knowledge structures themselves (via their work on the classification systems they then use). This is very different from the typical self-perception of AI researchers, who tend to see challenge #3 as mostly outside their sphere of influence and also their competence.

[17] Nunes, Zita Cristina: Remembering the Howard University Librarian Who Decolonized the Way Books Were Catalogued, in: Smithsonian Magazine, 26.10.2018. Online: <https://www.smithsonianmag.com/history/remembering-howard-university-librarian-who-decolonized-way-books-were-catalogued-180970890/>, Retrieved: 25.08.2022.

[18] Ibid.

As librarians are aware of these challenges, they are developing creative methods for dealing with them. For example, Leslie Howard, librarian at an elementary school in Virginia, observed that (as a consequence of choices such as those described above) while Thanksgiving, Christmas and Easter were catalogued under *390 / Social Sciences*, Eid, Rosh Hashanah and Diwali were catalogued under *290 / Other Religions*. "Howard's school has a diverse population, and she wants every student in the building who celebrates a religious holiday to find books about all of them in the same place. The pandemic gave her the flexibility to devote time to the project this year, since the district allowed her to tackle tasks not related to student achievement."[20] Howard re-arranged the books about religious holidays in her library accordingly. This solution was thus a super-imposition of a concrete spatial organisation (books next to one another for pupils to interact with) onto the abstract DDC organisation.

The algorithms employed in learning and cataloguing constitute challenge #2. Algorithms for natural language processing (i.e., for the processing of the full texts to be catalogued) increasingly rely on auxiliary structures such as word embeddings or pre-trained language models, which 'enrich' the information contained in the to-be-classified texts. In recent years, these have come under scrutiny for embedding the biases inherent in the large corpora they learn from (e.g., Bolukbasi et al., 2016; Caliskan, Bryson, & Narayanan, 2017), which may then influence the "downstream tasks" such as keyword classification (the analogue of subject indexing in the library context).[21]

We are not aware of any published work on the effects of word embeddings or pre-trained language models on keyword classification with library ontologies. To demonstrate how such effects could unfold, we created a fictitious example.[22] In her foundational AI textbook "Artificial Intelligence and Natural Man",[23] Margaret Boden used the simile that "computer programs are like knitting patterns". This was (and remains) an unusual comparison in the (still) male-dominated field of AI – even though it is actually historically motivated in the sense that the encoding of weaving patterns in punchcards and their use by mechanical looms constituted the first decoupling of "software" from "hardware" and in this sense "programming". Using Text Synth, an online interface for GPT-2 (the precursor to the current state-of-the-art text generation system GPT-3, Brown et al., 2020),[24] we showed how the sentence "Margaret Boden compares computer programs to knitting patterns" is enriched with text that is full of keywords from knitting/textiles/art rather than computer programs, which in turn might make a downstream classification algorithm catalogue the book under women's literature. Of course, any simile or metaphor could lead machine language processing astray. The point here is that pre-trained models will enrich wording based on statistics, i.e. based on associations that occur frequently, while they will recognise unusual descriptions not as context but as more "content" and enrich them in the same way as other content. What impact exactly pre-trained models have on the downstream tasks, and how such impact should be measured, are open research questions.[25] However, it is unlikely that biases that exist in the input data will have no effects; and the deployment of such models should be observed and monitored.

---

[19] Schwartz, Molly: Classifying books, classifying people, The Bytegeist Blog, 03.05.2018, <https://medium.com/the-bytegeist-blog/classifying-books-classifying-people-302430282a3d>.

[20] Joseph, Christina: Move Over, Melvil! Momentum Grows to Eliminate Bias and Racism in the 145-year-old Dewey Decimal System, in: School Library Journal, 18.08.2021. Online: <https://www.slj.com/story/move-over-melvil-momentum-grows-to-eliminate-bias-and-racism-in-the-145-year-old-dewey-decimal-system>, Retrieved: 25.08.2022.

[21] Bolukbasi, Tolga; Chang, Kai-Wei; Zou, James u. a.: Man is to Computer Programmer as Woman is to Homemaker? Debiasing Word Embeddings, in: 30th Conference on Neural Information Processing Systems (NIPS 2016), Barcelona, Spain 2016; Caliskan, Aylin; Bryson, Joanna J.; Narayanan, Arvind: Semantics derived automatically from language corpora contain human-like biases, in: Science 356 (6334), 14.04.2017, pp. 183–186. Online: <https://doi.org/10.1126/science.aal4230>.

[22] Berendt, Bettina: Unzulässige Diskriminierung durch KI-Systeme: (Nicht nur) technische Ursachen und Lösungsansätze, in: Conference vBIB21, Online 01.12.2021. Online: <https://people.cs.kuleuven.be/~bettina.berendt/Talks/berendt_2021_12_01.pdf >, Retrieved: 25.08.2022

[23] Boden, Margaret A.: Artificial intelligence and natural man, Hassocks 1977.

Challenge #3 concerns the classification systems themselves. Librarians have long observed a problem that computer-science ontology engineers describe as the differential "density" of taxonomies and other ontologies. For example, DDC differentiates its category *2 / Religions* into seven Christianity-related sub-categories (22–28) and groups all others into *29 / Other religions*. (This is complemented by the similarly coarse category *21 / Natural Theology and Secularism.*) This makes Christianity-related content more visible in a more nuanced way and in this sense, disadvantages other religions (see the observations on visibility, diversity and bias in Section 2 above).

A second question is what the position of a category in a taxonomy does to its perception. Parent-child relationships often carry hypernym-hyponym semantics, and sibling nodes can – even if this is not strictly intended or inherent in the logical semantics – confer an impression of similarity. Interpreted in this way, taxonomies reveal welcome changes in social perceptions but also persisting problems. For example, LGBTQI+ was classified in DDC in 1932 under 'Abnormal psychology' and under 'Social problems' in 1989. In 2018, "[y]ou'll find it at '306.7 —Sexual orientation, transgenderism, intersexuality'. To be clear, the current classification is an improvement but... I mean... LGBTQ+ people are [Groups of] PEOPLE (aka: 305), so finding LGBTQ+ related content "between prostitution and child trafficking on one side and fetishes and BDSM on the other" feels truly wrong. We can do better."[26]

Ontologies, as "formal, explicit specifications of shared conceptualisations", are models of conceptualisations that form the backbones of our natural languages and communication, but any such conceptualisation will embody values and normativity.[27] While they may be shared by many, they will seldom be shared by all and therefore be inherently contested. The question is how the normativities can change to reflect social progress – in the fluidity of everyday or (social) media language use, or in more binary, hierarchical, and instantaneous forms in explicitly coded ontologies.

---

[24] Brown, Tom; Mann, Benjamin; Ryder, Nick u. a.: Language Models are Few-Shot Learners, in: Larochelle, H.; Ranzato, M.; Hadsell, R. u. a. (Eds.): Advances in Neural Information Processing Systems, 2020, pp. 1877–1901. Online: <https://proceedings.neurips.cc/paper/2020/file/1457c0d6bfcb4967418bfb8ac142f64a-Paper.pdf>.

[25] Delobelle, Tokpo, Calders, u. a.: Measuring Fairness with Biased Rulers, 2022

[26] Laird, Katie: Something is rotten in the Dewey Decimal system, 2021, <https://www.careharder.com/blog/systemic-injustice-in-the-dewey-decimal-system>, Retrieved: 25.08.2022; O'Hara, Maria: Bad Dewey, Goldsmiths University of London library blog, 12.05.2021, <https://sites.gold.ac.uk/library-blog/bad-dewey/>, Retrieved: 25.08.2022.

The meaning and perception of a category in an ontology is influenced not only by its explicit neighbours in that ontology, but also by the manifold natural-language associations of the category label. Thus, naming is important, and naming does not automatically solve all problems. A recent change to the LCSH illustrates this:

> "[In November 2021,] the Library of Congress (LOC) made a change celebrated by a wide range of organizations, including the American Library Association. After years of pushing to make changes to the cataloging subject headings 'aliens' and 'illegal aliens,' the LOC replaced them with the terms 'noncitizen' and 'illegal immigration.' The decision has been discussed since at least 2016 when Congress's conservative politicians intervened and determined the headings would stay as-is.
>
> Media covering the change call it 'more accurate' and 'less offensive,' and the American Library Association said it was not only praiseworthy but that it 'better reflects common terminology and respects library users and library workers from all backgrounds. It also reflects the core value of social justice for ALA members.'
>
> While it certainly feels like progress to remove the term 'alien,' the problem rests in the fact that people are still being referred to as 'illegal'."[28]

While the story of Dorothy Porter highlights the role of individuals in the library profession as agents for change, in this case a campaign of many individuals (many of whom are conceivably not library professionals), indexed by the hashtag "#DropTheIWord", lobbied for the change.[29] This raises the question of who gets to define what is considered biased, unfair, or discriminatory, and how relevant stakeholders can be involved in processes of category evolution. We will return to this question under challenge #5 below.

## 3.2 Challenges for content analysis in direct user interaction

Modern content extraction techniques are less monolithic than keyword assignment. In particular, they are much more user-centric and thus dynamic. However, this creates further challenges.

---

[27] Gruber, Thomas R.: Towards principles for the design of ontologies used for knowledge sharing. in N. Guarino and R. Poli (Eds.), Formal Ontology in Conceptual Analysis and Knowledge Representation. Kluwer Academic Publishers, Deventer, The Netherlands, 1993.
[28] Jensen, Kelly: Library of Congress Subject Heading Change Doesn't Address The Real Issue, Book Riot, 15.11.2121, <https://bookriot.com/library-of-congress-subject-heading-change/>.
[29] Ibid.

Challenge #4 arises when algorithms learn from user interactions. Biased, unfair, and discriminatory speech can be generated inadvertently. For example, query-completion algorithms that are based on popularity can lead to people receiving sexist, racist, etc. claims as suggestions for what they "may be looking for". Thus, UN Women found that in 2013, Google recommended, as a completion to "women shouldn't …", "… have rights", "… vote", or "… work".[30] After these observations, the major Web search engines have restricted (probably largely manually) completion suggestions for many potentially offensive query seeds or combinations. However, such approaches are necessarily brittle, since in principle what is required is human-level natural-language understanding and world knowledge. Even in late 2021, "vaccination creates ..." produced the suggestion "… mutations" on German Google.[31] In sum, anyone offering query completions bears a two-fold responsibility because vulnerable individuals and groups may be harmed by (a) the representation-related harms of prejudices about them being made salient through suggestions and (b) the allocation-related harms resulting from following the suggestions and then receiving inferior-quality information. Many librarians are very aware of these risks and their responsibility towards their users, and this awareness may be one reason why query suggestions are not a popular interface choice for library catalogues today. Another reason may be that few such systems are available – but given the technological ease with which existing query-suggestion algorithms could be applied in library-oriented software, it may be argued that this lack of research and production is in itself a consequence of a lack of interest by librarians.

Challenge #5 is then how to integrate users – from librarians via library users to wider publics – in different ways in content extraction. To some extent, this is an attempt at a solution to challenge #3, or at the very least a channelling of the creative energy of users. This can take place within the traditional structure of having and using a classification system. For example, in recent years DDC has opened up to participatory design. It is increasingly emphasising the "local options", stating that "[t]o be clear, libraries using Dewey are always welcome to implement local solutions to areas where standard notation does not serve their users well."[32]

A proposal that goes beyond this is an idea for improving the static indexing of the Visual Holocaust Archive's interview videos (Presner 2016).[33] A first observation illustrates the power of drawing on the practices of cultures and academic fields: "the process [of assigning index terms] is exactly the opposite of what historians usually do, namely to create narratives from data by employing source material, evidence, and established facts into a narrative." Presner then proceeds to propose new practices: "we might imagine how a fluid data ontology might work, by allowing multiple thesauruses that recognize a range of knowledge, standards, and listening practices [...] What if a more participatory architecture allowed for other listeners to create tags that could unsay the said, or in other words, undo – or, at least, supplement – the definitive indexing categories and keywords associated with the segmented testimonies? […] Such structures of saying and unsaying the database would constantly re-interpret and re-inscribe the survivors' stories in ways that not only place the listener into an active relationship of responsibility but unleash a potentiality of meaning in every act of 'saying' and 'browsing'."[34] While some elements of these ideas may be said to be implemented even in standard folksonomy tagging systems, Presner's perspective opens the space for a wider discussion of responsibilities and interactions between libraries and users.[35]

---

[30] UN Women ad series reveals widespread sexism, UN Women, 21.10.2013, <https://www.unwomen.org/en/news/stories/2013/10/women-should-ads>, Retrieved: 28.08.2022.

[31] "impfen erzeugt → mutationen"

[32] Fox, Violet: Local option for classifying works about Indigenous peoples, 025.431: The Dewey blog, 10.01.2019, <https://ddc.typepad.com/025431/2019/01/local-option-for-classifying-works-about-indigenous-peoples.html>, Retrieved: 28.08.2022.

[33] Presner, Todd: The Ethics of the Algorithm: Close and Distant Listening to the Shoah Foundation Visual History Archive, in: Kansteiner, Wulf; Presner, Todd; Fogu, Claudio (Eds.): Probing the Ethics of Holocaust Culture, 2016, pp. 167–202. Online: <https://doi.org/10.4159/9780674973244-009>, Retrieved: 25.08.2022.

## 4 Conclusions: diversity interventions for fairness – and beyond

We have described several challenges and countermeasures arising in library contexts through a lens of fairness. But they could also be described through a lens of diversity. For example, Presner's proposal increases variety of the people whose perceptions are manifested in the metadata (and it may have effects on the other components of diversity in the metadata), and Dorothy Porter made black authors more visible by increasing variety, balance and disparity in categories such as poetry. Howard's re-organisation of books likewise challenged disparity by bringing seemingly disparate books physically close together, and changes in ontologies alter disparity (in the sense of ontological distance) between categories or between categories and natural-language associations.

How can AI and natural language processing help towards this goal? How can we leverage a fine-grained analysis of natural language texts to detect the diversity present in a given text? Since many users of such a system not only manage existing texts but also create new ones themselves, can awareness of the diversity present in a given text help them increase the diversity in texts they create? To answer these questions in the field of journalism, we have created the Diversity Searcher tool.[36] We hope that the observations described in this article and the use of the tool in libraries that we are currently starting, can form the basis for a constructive discussion between the communities of media and journalism studies, AI / natural language processing, and libraries.

---

[34] Ibid.

[35] Peters, Isabella; Kipp, Margaret E. I.; Heck, Tamara u. a.: Social tagging & folksonomies: Indexing, retrieving… and beyond?, in: Proceedings of the American Society for Information Science and Technology 48 (1), 2011, pp. 1–4. Online: <https://doi.org/10.1002/meet.2011.14504801069>.

[36] Berendt, Bettina; Karadeniz, Özgür; Mertens, Stefan and d'Haenens, Leen: Fairness beyond "equal": The Diversity Searcher as a Tool to Detect and Enhance the Representation of Socio-political Actors in News Media, in: Companion Proceedings of the Web Conference 2021, Ljubljana Slovenia 2021, pp. 202–212; Berendt, Bettina; Karadeniz, Özgür; Kıyak, Sercan; Mertens, Stefan and d'Haenens, Leen: Diversity and bias in DBpedia and Wikidata as a challenge for text-analysis tools. 2023. To appear in *o-bib. Das offene Bibliotheks-Journal.* Author version at https://people.cs.kuleuven.be/~bettina.berendt/DIAMOND/

# Acknowledgements

We thank the Fonds Wetenschappelijk Onderzoek – Vlaanderen (FWO) for funding DIAMOND (Diversity and Information Media: New Tools for a Multifaceted Public Debate) under project code S008817N.